\documentclass[12pt,a4paper]{article}
\usepackage{mathtools}
\usepackage{mathrsfs}
\usepackage{amsmath}
\usepackage{tikz-feynman}
\usepackage[utf8x]{inputenc}
\usepackage{amssymb}
\usepackage{slashed}
\usepackage{bbm}
\usepackage{bm}
\usepackage{color}
\usepackage[a4paper,top=3cm,bottom=3cm,left=2cm,right=3cm,bindingoffset=5mm]{geometry}
\usepackage{booktabs} 
\usepackage{tikz}
\usepackage{cancel}
\usepackage[bottom]{footmisc}
\usepackage{cite}
\usepackage{float}
\usepackage{todonotes}
\usepackage{jheppub}
\usepackage{bbold}

\usetikzlibrary{shapes.misc}
\tikzset{
    point/.style={
    draw=black,
    cross out,
    inner sep=0pt,
    minimum width=4pt,
    minimum height=4pt,
    },
}

\usepackage{hyperref}

\hypersetup{
colorlinks=true,         
linkcolor=blue,          
citecolor=red,        
urlcolor=cyan            
}

\renewcommand{\[}{\begin{equation}\begin{aligned}}
\renewcommand{\]}{\end{aligned}\end{equation}}

\newcommand{\wb}{{\bar{w}}}
\newcommand{\Pd}{\buildrel{\leftrightarrow} \over {P}}
\newcommand{\cev}[1]{\reflectbox{\ensuremath{\vec{\reflectbox{\ensuremath{#1}}}}}}

\title{Anomaly and double copy in quantum self-dual Yang-Mills and gravity}
\author{Ricardo Monteiro,}
\author{Ricardo Stark-Much\~ao,}
\author{Sam Wikeley}

\affiliation{Centre for Theoretical Physics, Department of Physics and Astronomy, \\ Queen Mary University of London, E1 4NS, United Kingdom}

\emailAdd{ricardo.monteiro@qmul.ac.uk}
\emailAdd{r.j.stark-muchao@qmul.ac.uk}
\emailAdd{s.wikeley@qmul.ac.uk}

\abstract{
Recent works have explored how scattering amplitudes in quantum self-dual Yang-Mills theory and self-dual gravity can be interpreted as resulting from an anomaly, as first proposed by W.~Bardeen. We study this problem in the light-cone-gauge formulation of the theories. Firstly, we describe how the infinite tower of symmetries associated to classical integrability can be quantum corrected, exhibiting the one-loop anomaly. Secondly, we present quantum-corrected light-cone Lagrangians worthy of the simplicity of the amplitudes, building on recent works describing the anomaly in twistor space. Finally, we discover an unexpected BCJ-like double copy for the (loop-integrated) amplitudes, distinct from the well-known BCJ double copy for the loop integrands.
}

\begin{document}

\begin{flushright}
QMUL-PH-22-36
\end{flushright}

\maketitle


\section{Introduction}
\label{sec:intro}

Self-dual Yang-Mills theory (SDYM) and self-dual gravity (SDG) are possibly the simplest four-dimensional field theories with non-trivial S-matrices. The only non-trivial contribution to the S-matrices is at one loop \cite{Bern:1993qk,Mahlon:1993si,Bern:1996ja,Bern:1998xc,Bern:1998sv}. Moreover, these one-loop amplitudes are very special in that they are rational functions of the kinematics of the external particles. It should be noted that SDYM / SDG are well-defined helicity sectors of full Yang-Mills theory / general relativity; see e.g.~\cite{Cangemi:1996rx,Chalmers:1996rq}.

Underlying the special features of quantum SDYM and SDG, somehow, is the integrability of the classical theories. At tree level, the standard argument is that integrability is associated to an infinite tower of symmetries, and this trivialises the tree-level S-matrix because no amplitudes could obey all such symmetries. But the classical integrability has also been argued -- originally by W. Bardeen \cite{Bardeen:1995gk} -- to underlie the simplicity of the one-loop amplitudes, which are supposed to result from the anomaly of the classical symmetries. Recent works \cite{Costello:2021bah,Costello:2022wso,Costello:2022upu,Bittleston:2022nfr,Bu:2022dis,Costello:2022jpg,Bittleston:2022jeq} (related to parallel developments in celestial holography \cite{Guevara:2021abz,Strominger:2021lvk,Jiang:2021ovh,Jiang:2021csc,Adamo:2021lrv,Ball:2021tmb,Mago:2021wje,Ren:2022sws,Monteiro:2022lwm,Bu:2022iak,Bhardwaj:2022anh,Guevara:2022qnm,Ball:2022bgg})
have provided an explicit realisation of this idea by uplifting the theories to twistor space, which is a natural framework for self-dual theories (see e.g.~\cite{Adamo:2017qyl,lionel1996integrability,Dunajski:2010zz}). The anomaly is identified as an obstruction to this uplift at one loop. A Green-Schwarz-type anomaly cancellation mechanism overcomes the obstruction via the inclusion of an `axion': the tree diagrams with axion exchanges cancel the loop diagrams. Effectively, this gives manifestly rational Feynman rules for the one-loop amplitudes in SDYM and SDG. 

Prior to the recent twistorial progress, Bardeen's suggestion motivated many investigations, e.g.~\cite{Rosly:1996vr,Brandhuber:2006bf,Brandhuber:2007vm,Boels:2008ef,Krasnov:2016emc,Nandan:2018ody,Chattopadhyay:2020oxe,Chattopadhyay:2021udc}. Several of these works made use of light-cone gauge, which is a natural choice in a spacetime approach, particularly for the self-dual theories. Manifest Lorentz symmetry is broken, but for a good cause: one gets a ghost-free action involving only the propagating degrees of freedom, and in the self-dual sector interactions are given by a simple cubic vertex.

In this work, we will study the quantum-corrected light-cone-gauge formulation of SDYM and SDG, building on the well-known light-cone actions \cite{Chalmers:1996rq,Siegel:1992wd}. By quantum corrected, we mean the inclusion of one-loop effective vertices {\it after} loop integration, instead of working with loop integrands as in previous works (see \cite{Lee:2022aiu,Gomez:2022dzk,Kakkad:2022ryl} for recent examples). Explicit versions of such effective vertices -- and manifestly rational ones -- are provided by the twistorial works mentioned above. We will consider these vertices in light-cone gauge, which will allow us to extend the already known vertices where needed -- in particular, for SU($N$) SDYM. The great surprise is that (a subset of) the effective vertices of SDYM `double copies' into the complete set of the effective vertices of SDG. This double copy takes a similar form as the BCJ colour-kinematics duality prescription \cite{Bern:2008qj,Bern:2019prr}, which is particularly simple (and valid off-shell) for the self-dual theories \cite{Monteiro:2011pc}. At loop-level, the BCJ prescription is aimed at the loop integrand \cite{Bern:2010ue}, and again that can be made manifest off-shell for the self-dual theories \cite{Boels:2013bi}. Here, unexpectedly, we find that a form of the double copy applies to the (loop-integrated) amplitudes. 

This paper is organised as follows. We discuss the basics of SDYM and SDG in light-cone gauge in section~\ref{sec:SD}, including aspects of classical integrability. In section~\ref{sec:qcformulation}, we discuss the structure of the light-cone-gauge quantum-corrected action and the appearance of the anomaly in the classical integrability currents. In section~\ref{sec:explicit}, we describe explicit constructions of the quantum-corrected actions for SDYM and SDG, expanding on recent results in the literature. We present in that section the unexpected double copy at the level of the (post-loop-integration) amplitude. Finally, we present a brief final discussion in section~\ref{sec:conclusion}.


\section{Basics of SDYM and SDG}
\label{sec:SD}

Here, we briefly review some basic features of self-dual Yang-Mills theory (SDYM) and self-dual gravity (SDG), particularly the light-cone formulations, the double-copy relation and the structure of classical integrability.

\subsection{Formulation in light-cone gauge}
\label{sec:formulation}

We will work with light-cone (in fact, double-null) coordinates $x^\mu=\{u,v,w,\wb\}$, such that the metric and the wave operator are
\[
ds^2=2(-du dv+dw d\wb)\,,\qquad \square=2(-\partial_u\partial_v+\partial_w\partial_\wb)\,.
\]

Let us start with SDYM. The action can be taken to be
\[
\label{eq:actionSDYM}
S_\text{SDYM}(B,A) = \int \text{tr}( B \wedge F_\text{ASD} )\,,
\]
where $B$ is a Lie-algebra-valued anti-self-dual 2-form, and $F_\text{ASD}$ is the anti-self-dual part of the Yang-Mills field strength $F$. We work in Minkowski spacetime, with the Hodge dual satisfying $\star^2=-1$; an anti-self-dual 2-form obeys $\star B = -iB$. 
In the action, $B$ acts as a Lagrange multiplier enforcing the self-duality of $F$. In Minkowski spacetime, a self-dual gauge field is necessarily complex, while split signature (2,2) admits real self-dual fields. We adopt the light-cone gauge by imposing
\[
\label{eq:Au}
A_u=0\,.
\]
Since $B$ is anti-self-dual, it has three independent components. Integrating out two of them enforces
\[
\label{eq:Awdd}
A_w=0 \qquad \text{and} \qquad \partial_u A_v = \partial_w A_\wb \,.
\]
The second equation is an integrability condition implying
\[
\label{eq:resAPhi}
A_v = \frac{1}{2}\partial_w \Psi\,, \qquad  A_\wb=\frac{1}{2}\partial_u \Psi \,,
\]
where we chose the numerical coefficient.
We are left with the action \cite{Chalmers:1996rq} (see also \cite{PARK1990287,Cangemi:1996rx,Bardeen:1995gk})
\[
\label{eq:SDYM}
S_\text{SDYM}(\Psi,\bar\Psi) = \int d^4 x \;\; \text{tr}\; \bar\Psi \big(\square \Psi + i[\partial_u \Psi,\partial_w \Psi]\big) \,,
\]
where $\bar\Psi$ is the remaining component of $B$. The fields $\Psi$ and $\bar\Psi$ should be thought of as describing positive and negative helicity degrees of freedom, respectively.

For SDG, the intermediate steps are more involved, but the story is similar. The steps analogous to \eqref{eq:Au}-\eqref{eq:resAPhi} lead to
\[
\label{eq:reshphi}
ds^2=2(-du dv+dw d\wb) + \partial_w^2\phi\,dv^2 + \partial_u^2\phi\,d\wb^2 +2 \partial_u\partial_w\phi\,dvd\wb
\,.
\]
One obtains the action \cite{Siegel:1992wd}
\[
\label{eq:SDG}
S_\text{SDG}(\phi,\bar\phi) = \int d^4 x \;\; \bar\phi \big(\square \phi - \{\partial_u \phi,\partial_w \phi\}\big)\,,
\] 
with the Poisson bracket
\[
\label{eq:Poisson}
\{f,g\}=\partial_u f \,\partial_w g -\partial_w f\, \partial_u g\,.
\]

The momentum-space Feynman rules for the actions \eqref{eq:SDYM} and \eqref{eq:SDG} are very simple. We follow a convention where all the particles in a vertex or diagram are taken to be incoming, which allows us to speak unambiguously about their helicities. We have the following Feynman rules.
\begin{itemize}
\item Propagator $(+-)$:\, $\frac{1}{k^2}\,\delta^{ab}$ in SDYM, and $\frac{1}{k^2}$ in SDG.
\item Cubic vertex $(++-)$:
\[ 
\label{eq:3ptvert}
V_\text{SDYM} = X(k_1,k_2) \,f^{a_1a_2a_3}\,, \qquad V_\text{SDG} = X(k_1,k_2)^2 \,,
\]
with
\[
\label{eq:X}
X(k_1,k_2)=k_{1w}k_{2u}-k_{1u}k_{2w}=-X(k_2,k_1) \,.
\]
The spinor-helicity formalism is very convenient for computations. In this language, we have
\[
\label{eq:Xsh}
X(k_1,k_2)=\langle \eta | k_1k_2 | \eta\rangle \,, 
\]
where $| \eta\rangle$ is the reference spinor of the light-cone gauge direction. We follow the notation of \cite{Boels:2013bi}.
\item External state factors: given, in the spinor-helicity formalism, by $\langle \eta i \rangle^{\mp 2s}$ for a $\pm$-helicity gluon ($s=1$) or graviton $s=2$.
\end{itemize}
Using these rules, the only diagrams that one can draw are the following.
\begin{itemize}
\item Tree level: $n$-point diagrams with $n-1$ positive helicity external particles, and one negative helicity external particle.
\item One loop: $n$-point diagrams with $n$ positive helicity external particles, and no negative helicity external particle.
\end{itemize}
These diagrams are precisely those that can be drawn in the full Yang-Mills theory or general relativity for these loop and helicity choices, so SDYM and SDG are well-defined sectors of the full theories. Unlike the full theory, though, it is not possible to draw any two- and higher-loop Feynman diagrams in the self-dual theories, which are therefore one-loop exact theories. In fact, the S-matrices are given by the one-loop amplitudes (non-trivial for $n\geq4$), because the tree-level amplitudes vanish due to classical integrability.\footnote{The tree-level three-point amplitude is a special case: three-point on-shell kinematics requires the complexification of the momenta (or analytic continuation to split signature). These amplitudes play a crucial role in modern on-shell methods, e.g.~\cite{Britto:2005fq}, due to a combination of locality$+$unitarity and Cauchy's residue theorem.} 

One important feature of SDYM is that it provides a rare off-shell example of colour-kinematics duality in gauge theory \cite{Bern:2008qj,Bern:2019prr}, as described in \cite{Monteiro:2011pc,Boels:2013bi}; see also \cite{Campiglia:2021srh}. The `duality' between the colour Lie algebra and a kinematic algebra is manifest in the SDYM vertex \eqref{eq:3ptvert}. While $f^{a_1a_2a_3}$ are the structure constants of the colour Lie algebra, which we take to be SU($N$), $X(k_1,k_2)$ are the structure constants of the Lie algebra of area-preserving (i.e.~unit-Jacobian) diffeomorphisms in the $u$-$w$ plane, generated by vector fields\, $L_{k}=\{e^{i\,k\cdot x},\cdot\}$\,. Then, the SDG vertex \eqref{eq:3ptvert} is the square of the kinematic part of the SDYM vertex, i.e., it exhibits two copies of the kinematic algebra. Beyond the self-dual sector, the colour-kinematics duality is a more intricate story; see e.g.~\cite{Cheung:2021zvb,Brandhuber:2021bsf,Chen:2022nei,Brandhuber:2022enp,Cao:2022vou} for recent studies, and \cite{Bern:2019prr,Bern:2022wqg,Kosower:2022yvp,Adamo:2022dcm,Mafra:2022wml} for related reviews.

\subsection{Classical integrability}
\label{sec:classint}

We will focus on reviewing SDYM here, and there is an analogous story for SDG. The (classical) equations of motion are
\begin{align}
\label{eq:eomPsi}
& 0 
= \square \, \Psi+ i[\partial_u \Psi,\partial_w \Psi]  \,, \\
& 0 
= \square \, \bar \Psi+ i[\partial_u \bar \Psi,\partial_w \Psi] + i[\partial_u \Psi,\partial_w \bar \Psi] \,.
\label{eq:eomPsib}
\end{align}
The first equation is the self-duality condition on the gauge field, and holds also in the quantum-corrected theory, due to it being enforced in the action \eqref{eq:SDYM} via a Lagrange multiplier. Crucially, the second equation can be interpreted as a linearised deformation of a solution to the first equation, $\Psi\to \Psi+\epsilon \,\bar\Psi$. It can be expressed as the conservation of a current, $\partial_\mu { J}^\mu =0$,
 if we define
\begin{equation}
\label{eq:J}
{ J} = \left(\partial_\wb \bar\Psi + \tfrac{i}{2}[\partial_u \Psi, \bar\Psi]\right)\partial_w - \left(\partial_v \bar\Psi + \tfrac{i}{2}[\partial_w \Psi, \bar\Psi]\right)\partial_u \,.
\end{equation}

The integrability of the classical theory is revealed by the fact that there is an infinite tower of conserved currents,
\begin{equation}
\label{eq:Jr}
{ J}_r = { J}(\bar\Psi=\Lambda_r) \,, \quad r=0,1,2,\cdots,
\end{equation}
that is, there is an infinite tower $\{\Lambda_r\}$ of solutions to \eqref{eq:eomPsib}. Each solution can be interpreted as a linearised deformation of a solution to \eqref{eq:eomPsi}, $\Psi\to\Psi+\epsilon \Lambda_{r}$. The tower is formally constructed as follows \cite{Prasad:1979zc,Dolan:1983bp}: one starts with a seed solution $\Lambda_0$, and then recursively obtains the elements $\Lambda_r$ for $r>0$ using the relations
\begin{equation}
\partial_u\Lambda_{r+1} =  \partial_\wb \Lambda_{r} + \tfrac{i}{2}[\partial_u \Psi,\Lambda_{r}] \,, \qquad
\partial_w\Lambda_{r+1} =  \partial_v \Lambda_{r} + \tfrac{i}{2}[\partial_w \Psi,\Lambda_{r}] \,,
\label{eq:pairrelations}
\end{equation}
equivalent to $\{\Lambda_{r+1},\cdot\}={ J}_{r}$. These relations at level $r+1$ are compatible because $\bar\Psi=\Lambda_{r}$ is a solution to \eqref{eq:eomPsib}:
\begin{equation}
\partial_w(\partial_u\Lambda_{r+1})-\partial_u(\partial_w\Lambda_{r+1}) = \tfrac{i}{2}\big(\square \, \Lambda_{r}+ i[\partial_u \Lambda_{r},\partial_w \Psi] + i[\partial_u \Psi,\partial_w \Lambda_{r}] \big)=0\,.
\end{equation}
It then follows that $\bar\Psi=\Lambda_{r+1}$ is also a solution to \eqref{eq:eomPsib}, by virtue of \eqref{eq:eomPsi},
\begin{equation}
\square \, \Lambda_{r+1}+ i[\partial_u \Lambda_{r+1},\partial_w \Psi] + i[\partial_u \Psi,\partial_w \Lambda_{r+1}] = \tfrac{i}{2}[\square \, \Psi+  i[\partial_u \Psi,\partial_w \Psi],\Lambda_{r}]=0 \,.
\end{equation}
Hence, given a solution to \eqref{eq:eomPsib}, we can in principle move in the solution space spanned by the tower $\{\Lambda_r\}$.

The integrability is also commonly expressed in terms of a Lax pair, which in our gauge choice is given by
\[
\mathcal{L} = \partial_u - \lambda \left(\partial_\wb+\tfrac{i}{2}\partial_u\Psi\right)\,, \qquad 
\mathcal{M} = \partial_w - \lambda \left(\partial_v+\tfrac{i}{2}\partial_w\Psi\right)\,,
\]
where $\lambda \in \text{CP}^1$ is the spectral parameter. Then $[\mathcal{L},\mathcal{M}]=0$ reproduces equation \eqref{eq:eomPsi}. The tower discussed above arises here from imposing
\[
\mathcal{L}\, \Lambda = 0 = \mathcal{M}\, \Lambda \qquad \text{on} \qquad \Lambda(x,\lambda)=\sum_{r=0}^\infty \Lambda_r(x) \lambda^r\,.
\]

See ref.~\cite{lionel1996integrability} for an overview of the classical integrability aspects of SDYM, and the connection to twistor theory.

An implication of classical integrability is that there is no particle production at tree level, i.e.~the tree scattering amplitudes vanish. The intuitive notion is that no amplitude can be defined that obeys the infinite tower of symmetries. The vanishing of the amplitudes can be derived from the Feynman rules presented earlier. The proof follows by the inductive construction of the perturbative solutions to \eqref{eq:eomPsi} \cite{Bardeen:1995gk,Cangemi:1996rx}. 

The presence of the area-preserving diffeomorphism algebra is tightly connected to the classical integrability. In fact, all classically integrable systems are conjectured to be some `reduction' of SDYM \cite{Ward:1985gz}, which in the four-dimensional known cases simply means substituting $f^{a_1a_2a_3}$ in the cubic vertex by the structure constants of another Lie algebra. In terms of scattering amplitudes, this can be understood from the following: in any theory with a cubic vertex similarly exhibiting two algebras, if one is the kinematic algebra of area-preserving diffeomorphisms, then the tree-level amplitudes in the theory vanish by virtue of 
the double copy \cite{Monteiro:2022lwm} (see also \cite{Chacon:2020fmr}).


\section{General aspects of a quantum-corrected formulation}
\label{sec:qcformulation}

\subsection{Quantum action}

The actions \eqref{eq:SDYM} and \eqref{eq:SDG} define SDYM and SDG. The computation of perturbative quantum effects proceeds as usual from the Feynman rules with the consideration of diagrams with loops. In the case of the self-dual theories, only one-loop diagrams arise, as we discussed. The fact that the theories are one-loop exact means that we obtain a quantum-corrected action (i.e.~one for which quantum computations proceed in a classical-like manner) by simply introducing new vertices associated to one-loop polygon (sub)diagrams, i.e., off-shell diagrams with legs directly attached to the loop. Moreover, the Feynman rules for the self-dual theories imply that all such one-loop vertices have only positive-helicity legs attached. This leads to the general structure of the quantum-corrected SDYM action,
\[
\label{eq:qSDYM}
S_\text{q.c.SDYM}(\Psi,\bar\Psi) = \int d^4 x \;\Big( \text{tr}\; \bar\Psi \big(\square \Psi + i[\partial_u \Psi,\partial_w \Psi]\big) + V_{\text{1-loop}}[\Psi] \Big) \,.
\]
The new part $V_{\text{1-loop}}$, which is suppressed by a factor of $\hbar$, depends only on $\Psi$, not on $\bar\Psi$ (since it admits only positive-helicity legs attached), and is non-local. It contains all the one-loop vertices:
\[
V_{\text{1-loop}}[\Psi] = \sum_{m=2}^\infty V^{(m)}_{\text{1-loop}}[\Psi] \,,
\qquad \text{where} \quad V^{(m)}_{\text{1-loop}}[\Psi] \sim \Psi^m \,,
\]
Here, $V^{(2)}_{\text{1-loop}}[\Psi]$ comes from an off-shell bubble diagram, $V^{(3)}_{\text{1-loop}}[\Psi]$ from an off-shell triangle diagram, $V^{(4)}_{\text{1-loop}}[\Psi]$ from an off-shell box diagram, and so on. We are just explicitly inserting the loop sub-diagrams as vertices, so it is clear that the `tree' amplitudes of the quantum-corrected action give the complete amplitudes of the theory. Entirely analogous statements apply to SDG.

As in any off-shell formulation of the theories, field redefinitions can lead to dramatically different actions; in particular, $V_{\text{1-loop}}$ is non-unique. Since the scattering amplitudes in SDYM and SDG are rational functions of the external kinematic data, one may expect that there exist choices of $V_{\text{1-loop}}$ where these vertices are rational (albeit non-local) functions in momentum space. Let us consider three very distinct examples of choosing $V_{\text{1-loop}}$.
\begin{itemize}

\item {\it Momentum-space vertices arising from the anomaly cancellation in twistor space}, as achieved in \cite{Costello:2021bah,Costello:2022wso,Bittleston:2022nfr}. These have beautiful properties, among which (i) the vertices are rational functions of the external data, and (ii) the vertices $V^{(m)}_{\text{1-loop}}$ are non-trivial only for $m\geq4$, which matches the fact that the one-loop $n$-point amplitudes are non-trivial only for $n\geq4$. This will be our main choice in this paper, where we will partly generalise the SDYM results in \cite{Costello:2022wso} to SU($N$), and uncover a double-copy relation between the vertices in SDYM and SDG. 

\item {\it A single `region-momenta vertex'} \cite{Brandhuber:2007vm} {\it arising from a specific regularisation of the one-loop bubble diagram} \cite{Thorn:2004ie,Bardakci:2002xi,Chakrabarti:2005ny}; see also \cite{Boels:2007gv} and \cite{Chattopadhyay:2020oxe,Chattopadhyay:2021udc}. This procedure relies on planarity and applies to the SDYM case. On the one hand, it gives simple and manifestly rational rules for the computation of the one-loop amplitudes. On the other hand, the region-momenta do not directly translate into a spacetime approach, and the rules do not allow easily for the construction of perturbative solutions to the quantum equations of motion using the Berends-Giele recursion, because the region-momenta differ at each step in the recursion. We will not pursue this choice, though it would be interesting to understand better how it connects with other developments.

\item {\it Momentum-space vertices arising directly from the light-cone-gauge Feynman rules}, in particular \eqref{eq:3ptvert}, after explicitly performing the loop integration of off-shell bubble/triangle/box/etc diagrams using a standard regularisation. For illustration, this brute-force approach is described in appendix~\ref{sec:bruteforce} for SDYM, focusing on the first few values of $m$ in $V^{(m)}_{\text{1-loop}}[\Psi]$. The basic features are:  $V^{(2)}_{\text{1-loop}}[\Psi]=0$\,; $V^{(3)}_{\text{1-loop}}[\Psi]$ is non-vanishing (even though the three-point one-loop amplitudes vanish) and naively quite fearsome, involving dilogarithms, although a closer inspection reveals that a usable rational vertex can be obtained. For higher $m$, the computations become increasingly more challenging, but it appears that rational vertices can be obtained in this manner.

\end{itemize}

\subsection{Anomalous classical symmetries}

Let us again focus on SDYM, since the SDG case is similar. The quantum-corrected action \eqref{eq:qSDYM} leads to the quantum-corrected equations of motion:
\begin{align}
\label{eq:qeomPsi}
& 0
= \square \, \Psi+ i[\partial_u \Psi,\partial_w \Psi]  \,, \\
& 0
= \square \, \bar \Psi+ i[\partial_u \bar \Psi,\partial_w \Psi] + i[\partial_u \Psi,\partial_w \bar \Psi] +
\frac{\delta{V_{\text{1-loop}}[\Psi]}}{\delta \Psi} \,.
\label{eq:qeomPsib}
\end{align}
The first equation is unchanged from the classical case \eqref{eq:eomPsi}, as already stated. The second equation is affected by the quantum correction. Crucially, while the first equation is still integrable, the second equation is no longer a linearisation of the first equation around a background, so the classical integrability of the complete theory is broken at the quantum level.

One obvious question, first asked by Bardeen \cite{Bardeen:1995gk}, is what happens to the infinite tower of currents ${ J}_r$ associated to classical integrability. If these currents are defined based only on equation \eqref{eq:qeomPsi} and its linearisation, which are not quantum corrected, then the currents are also conserved in the quantum theory. That is, the currents ${ J}_r$ defined via $J$ as in \eqref{eq:Jr} are still conserved. However, $J$ itself, as defined in \eqref{eq:J}, is now not conserved because the quantum correction in \eqref{eq:qeomPsib} results in
\begin{equation}
\partial_\mu {{ J}}^\mu =\frac1{2} (\square \, \bar \Psi+ i[\partial_u \bar \Psi,\partial_w \Psi] + i[\partial_u \Psi,\partial_w \bar \Psi]) =-\frac1{2}\, \frac{\delta{V_{\text{1-loop}}[\Psi]}}{\delta \Psi} \,.
\end{equation}
In order to obtain a conserved quantum-corrected version, let us notice first that the quantum-corrected action is naturally chosen to depend on $\Psi$ only through its derivatives, and in particular its $u$ and $w$ derivatives; indeed, that is already the case for the non-corrected action \eqref{eq:SDYM}. We will see explicit examples of quantum-corrected actions later. Hence, it follows from the variational integration by parts that
\[
\frac{\delta{V_{\text{1-loop}}[\Psi]}}{\delta \Psi} = \partial_u {\mathcal V}^u[\Psi]+ \partial_w {\mathcal V}^w[\Psi]\,,
\]
where ${\mathcal V}^u$ and ${\mathcal V}^w$ are (minus) the coefficients of the variations of $\partial_u\Psi$ and $\partial_w\Psi$, respectively; this splitting is not unique because there are appearances of $\partial_u\partial_w\Psi$, but it is possible. This leads to the quantum conserved current
\begin{equation}
\label{eq:qJ}
{ J}^\text{q.c.} =  \left(\partial_\wb \bar\Psi + \tfrac{i}{2}[\partial_u \Psi, \bar\Psi] + \tfrac1{2}\,{\mathcal V}^w\right)\partial_w - \left(\partial_v \bar\Psi + \tfrac{i}{2}[\partial_w \Psi, \bar\Psi] - \tfrac1{2}\,{\mathcal V}^u \right)\partial_u
\,.
\end{equation}

In analogy with \eqref{eq:Jr}, this suggests that we define
\begin{equation}
\label{eq:qJr}
{ J}_r^{\text{q.c.}} = { J}^{\text{q.c.}}(\bar\Psi=\Lambda_r) \,, \quad r=0,1,2,\cdots.
\end{equation}
By construction, these currents have a universal anomaly:
\begin{equation}
\label{eq:DJiq}
\partial^\mu {{ J}_r^{\text{q.c.}}}\!\!\!{}_\mu =\frac1{2}\, \frac{\delta{V_{\text{1-loop}}[\Psi]}}{\delta \Psi} \,, \quad \forall r\,.
\end{equation}
To summarise, the breaking of integrability at the quantum level suggests that the tower of classically conserved currents should be quantum-corrected so that those currents are no longer conserved, and this is should be tied up with the quantum-corrected equation of motion \eqref{eq:qeomPsib}. We have made here a proposal for how this works. The anomaly can be thought of as generating the one-loop amplitudes. The $m$-point one-loop vertices are associated to $\frac{\delta^m}{(\delta \Psi)^m}V_{\text{1-loop}}\Big|_{\Psi=0}=2\frac{\delta^{m-1}}{(\delta \Psi)^{m-1}}\partial^\mu {{ J}_r^{\text{q.c.}}}\!\!\!{}_\mu\Big|_{\Psi=0}$, and the amplitudes are obtained as a sum over these vertices dressed with trees.

It would be interesting to relate this proposed definition of the anomalous integrability currents to a twistor-based construction. Notice, however, that our definition is not unique, because $V_{\text{1-loop}}$ is not unique.

\subsection{Off-shell amplitudes as vertices}

There is another approach to the quantum action, which eliminates the `tree' vertex $\bar\Psi\Psi\Psi$ in \eqref{eq:qSDYM} in favour of turning the one-loop `off-shell amplitudes' directly into vertices. Following \cite{Brandhuber:2006bf,Feng:2006yy}, we perform a field redefinition where we write $\Psi$ as a function of $\tilde \Psi$, by perturbatively solving
\[
\square \Psi + i[\partial_u \Psi,\partial_w \Psi] = \square \tilde \Psi\,.
\]
This redefinition was introduced in the context of the tree-level MHV rules in the full Yang-Mills theory \cite{Cachazo:2004kj,Mansfield:2005yd}. For SDYM, we obtain
\[
\label{eq:tqSDYM}
S_\text{q.c.SDYM}(\tilde\Psi,\bar\Psi) = \int d^4 x \;\big(\, \text{tr}\; \bar\Psi \square \tilde\Psi  + \tilde V_{\text{1-loop}}[\tilde \Psi] \,\big) \,,
\]
where $\tilde V_{\text{1-loop}}[\tilde \Psi]= V_{\text{1-loop}}[ \Psi(\tilde\Psi)]$\,. Effectively, we are dressing the previous vertices with trees, so that the  $n$-point amplitude is obtained directly from the new $n$-point vertex $\tilde V_{\text{1-loop}}^{(n)}[\tilde \Psi]\sim \tilde \Psi^n$\,.\footnote{The amplitude is the on-shell vertex times the external helicity factors discussed in section~\ref{sec:SD}.} In this approach, the classical theory is trivialised. While this brutal `integrable$=$free' redefinition of the classical theory is not generally well defined, the new action is perfectly valid for the computation of scattering amplitudes (with real kinematics in Minkowski spacetime, so as to avoid the tree-level three-point amplitude). The analogue of the conserved current \eqref{eq:qJ} is simply
\begin{equation}
\label{eq:tqJ}
\tilde { J}_\text{q.c.} = ( \partial_\wb \bar\Psi + \tfrac1{2}\,\tilde{\mathcal V}^w)\, \partial_w - (\partial_v \bar\Psi - \tfrac1{2}\,\tilde{\mathcal V}^u)\,  \partial_u
\,.
\end{equation}
In the following, we will consider actions of the type \eqref{eq:qSDYM}, rather than \eqref{eq:tqSDYM}.


\section{Explicit quantum actions}
\label{sec:explicit}

In this section, we will present explicit quantum-corrected actions for SDYM and SDG, building on the works \cite{Costello:2021bah,Costello:2022wso,Bittleston:2022nfr}. In these works, modifications of SDYM (for restricted gauge groups) and of SDG with vanishing loop amplitudes are presented that can be lifted to twistor space at quantum level. The extension is the inclusion of an `axion' that couples to $F\wedge F$ in SDYM or to $R\wedge R$ in SDG, realising a Green-Schwartz-type anomaly cancellation whereby the tree exchanges involving the axion cancel the loop diagrams. Building on these works, we can obtain quantum-corrected actions for SDYM and SDG by (i) integrating out the axion exchange to obtain non-local effective vertices, and (ii) flipping the sign of the vertices. The sign flip is consistent with the vanishing of the amplitude in the modified models of \cite{Costello:2021bah,Costello:2022wso,Bittleston:2022nfr}. In the SDYM-axion model of \cite{Costello:2021bah}, we have
\[
\text{loop diagrams with gluon in loop} + \text{tree diagrams with axion exchange} = 0 \,, \nonumber \\
\]
hence
\[
\text{loop diagrams with gluon in loop} =-\, \text{tree diagrams with axion exchange}. \nonumber
\]
In the SDG-axion model of \cite{Bittleston:2022nfr}, the coupling of the axion is such that this field can also run in the loop, so that
\begin{align}
\text{loop diagrams with graviton in loop} = \; & -  \text{loop diagrams with axion in loop}   \nonumber \\
& - \text{tree diagrams with axion exchange}  \nonumber \\
= \; & - \frac1{2}\; \text{tree diagrams with axion exchange}  \,. \nonumber 
\end{align}
The last step follows from the fact the axion has a fourth-derivative kinetic term, and therefore its contribution running in the loop is equivalent to the contribution of the graviton in the loop \cite{Bittleston:2022nfr}.\footnote{This point was corrected in the current version of our paper.}
We will now see how this works in practice.

\subsection{Quantum Feynman rules for SU($N$) SDYM}

In \cite{Costello:2021bah,Costello:2022wso}, the following theory was described,
\[
\label{eq:Saxion}
S_{\rho-\text{SDYM}}(B,A,\rho) = \int \text{tr}\Big( B \wedge F_\text{ASD}\, +d^4 x \,\frac1{2}(\square\rho)^2 + \tilde a\,\rho\, F\wedge F\Big)\,,
\]
which is a modification of the action \eqref{eq:actionSDYM} for SDYM to include $\rho$, called axion because of how it couples to the gauge field; $\tilde a$ is a coupling constant. The cancellation mechanism between loop contributions of the gauge field and tree exchanges of the axion restricts the colour gauge group to be SU(2), SU(3), SO(8) or one the exceptional groups; for each of these choices the coupling $\tilde a$ takes a particular value that ensures the cancellation.

Following the reasoning at the beginning of the section, this implies that the quantum action of SDYM for those particular gauge groups (hence the prime $S'$) can be taken to be\footnote{Notice that integrating out the axion in \eqref{eq:Saxion} gives the opposite sign for the second term, so the sign flip discussed above provides the positive sign. Note also the two $\frac1{2}$ factors from the definition of the 2-form field strength.}
\[
S'_{\text{q.c.SDYM}}(B,A) = \int \text{tr}( B \wedge F_\text{ASD}) + \,d^4 x\,\frac{{\tilde a}^2}{32} \Big(\, \frac1{\square} \,\text{tr}(\varepsilon^{\mu\nu\rho\lambda}F_{\mu\nu}F_{\rho\lambda}) \Big)^2 \,.
\]
The steps towards obtaining a light-cone-gauge version of this action are exactly the same as in \eqref{eq:Au}-\eqref{eq:resAPhi}, because integrating out components of B is not affected by the non-local term. The result is simply
\[
\label{eq:qcSDYM'}
S'_{\text{q.c.SDYM}}(\Psi,\bar\Psi) = \int d^4 x \;\; \text{tr}\big( \bar\Psi \big(\square \Psi + i[\partial_u \Psi,\partial_w \Psi])\big) + a\, \Big(\, \frac1{\square} \,\text{tr}( \Psi \Pd\!{}^2\, \Psi ) \Big)^2 \,.
\]
The coupling $a$ is $\tilde a^2$ up to a numerical factor, which is fixed by the normalisation of the one-loop amplitude. Notice that $a$ is proportional to $\hbar$. We used the differential operator
\[
\label{eq:PoissonP}
\buildrel{\leftrightarrow} \over {P} \; = \;
\buildrel{\leftarrow} \over{\partial}_u
\buildrel{\rightarrow} \over{\partial}_w-
\buildrel{\leftarrow} \over{\partial}_w
\buildrel{\rightarrow} \over{\partial}_u\,,
\]
which in momentum space turns into the kinematic structure constant \eqref{eq:X}. This action is of the form anticipated in \eqref{eq:qSDYM}, since $V_{\text{1-loop}}$ depends on $\Psi$, not $\bar\Psi$. However, it only has a four-point `one-loop vertex', and this is where the restriction of the gauge group operates. Under this restriction, there are identities among colour traces that significantly simplify the amplitudes beyond four points.\footnote{This is used in the step from (11.0.7) to (11.0.8) in \cite{Costello:2022wso}.} Still, let us consider the Feynman rules arising from \eqref{eq:qcSDYM'}, where we will define our normalisation of the one-loop amplitudes by setting $a=1$.
\begin{itemize}
\item Propagator $(+-)$:\, $\frac{1}{k^2}\,\delta^{a_1a_2}$\,.
\item $\hbar^0$-vertex $(++-)$:\, $X(k_1,k_2)\,f^{a_1a_2a_3}$\,, with $X(k_1,k_2)=\langle \eta | k_1k_2 | \eta\rangle$\,.
\item  $\hbar^1$-vertex $(++++)$:\, $\displaystyle X(k_1,k_2)^2\,\frac1{s_{12}^2}\,X(k_3,k_4)^2\,\delta^{a_1a_2}\delta^{a_3a_4}$\,, \,with $s_{12}=(k_1+k_2)^2$\,.
\item External state factors: $\langle \eta i \rangle^{\mp 2}$ for a $\pm$-helicity gluon.
\end{itemize}
One-loop diagrams have a single one-loop vertex ($\hbar^1$), any remaining vertices being tree-level ones ($\hbar^0$). Recalling \eqref{eq:Xsh}, the four-point all-plus one-loop result follows from
\[
\label{eq:4ptCPcheck}
\Big(\prod_{i=1}^4\langle \eta i \rangle^{- 2}\Big)\,X(k_1,k_2)^2\,\frac1{s_{12}^2}\,X(k_3,k_4)^2 = \frac{[12]^2[34]^2}{s_{12}^2}=\frac{[12][34]}{\langle 12\rangle\langle 34\rangle}\,.
\]
This is the $s$-channel contribution, and there is also the $t$-channel contribution, which coincides with the former because the on-shell expression above turns out to be permutation invariant in the four particles.

One may proceed to calculate the one-loop higher-point amplitudes using the rules above, but a puzzle arises: they are not gauge invariant beyond four points, because they depend on the reference spinor $|\eta\rangle$. The solution to the puzzle is precisely the restriction of the colour group: the colour structures that are taken to be independent for SU($N$) are not independent in this case; so while the full (colour-dressed) amplitude is gauge-invariant, the naive colour-ordered amplitudes are not.

In order to extend the Feynman rules above to SU($N$), additional vertices must be included. In fact, there are two classes of vertices: ones that extend the four-point vertex above, and additional vertices carrying the permutation symbol. The SU($N$) colour-ordered amplitudes, conjectured in \cite{Bern:1993qk} and proven in \cite{Mahlon:1993si}, were already in \cite{Bern:1993qk} observed to split in such a manner:
\begin{align}
A^{(1)}_\text{SDYM}(123\cdots n) 
& = M_n  \sum_{1\leq i_1<i_2<i_3<i_4\leq n} \frac{\langle i_1i_2\rangle [i_2i_3]\langle i_3i_4 \rangle [i_4i_1] }{\langle 12\rangle\langle 23\rangle \cdots \langle n1\rangle}
 \nonumber \\
& = M_n \;\frac{E(123\cdots n)+O(123\cdots n)}{\langle 12\rangle\langle 23\rangle \cdots \langle n1\rangle}\,,
\end{align}
where $M_n$ is a numerical normalisation constant, while
\[
E(123\cdots n) = \sum_{1\leq i_1<i_2<i_3<i_4\leq n} \langle i_1i_2\rangle [i_2i_3]\langle i_3i_4 \rangle [i_4i_1] + [i_1i_2]\langle i_2i_3\rangle[i_3i_4]\langle i_4i_1\rangle
\]
and
\begin{align}
O(123\cdots n) &= \sum_{1\leq i_1<i_2<i_3<i_4\leq n} \langle i_1i_2\rangle [i_2i_3]\langle i_3i_4 \rangle [i_4i_1] - [i_1i_2]\langle i_2i_3\rangle[i_3i_4]\langle i_4i_1\rangle \nonumber \\
&= -\sum_{1\leq i_1<i_2<i_3<i_4\leq n} \varepsilon(i_1,i_2,i_3,i_4) = -\sum_{1\leq i_1<i_2<i_3<i_4\leq n-1} \varepsilon(i_1,i_2,i_3,i_4) \,,
\end{align}
with the last step following from momentum conservation. Notice that the $O$-part of the amplitude vanishes at four points. Notice also that the two parts arise from sets of vertices that are independent, because each contribution to the one-loop amplitude carries a single one-loop vertex. In the following, we will consider the colour-dressed amplitudes,
\[
{\mathcal A}^{(1)}_{n\;\text{SDYM}} = \sum_{\sigma\in S_{n-1}/{\mathcal R}} c^{a_{\sigma(1)}a_{\sigma(2)}a_{\sigma(3)}\cdots \, a_{\sigma(n)}} \,A^{(1)}_\text{SDYM}\big({\sigma(1)}{\sigma(2)}{\sigma(3)}\cdots {\sigma(n)}\big) 
\,,
\]
where the sum is over non-cyclic permutations, modulo reflection of the list $\sigma$. We employ the notation
\[
c^{a_1a_2a_3\cdots\, a_n} = f^{b_1a_1b_2}f^{b_2a_2b_3}f^{b_3a_3b_4}\cdots f^{b_na_nb_1}
\]
for the cyclic $n$-gon colour factors. It will be useful to denote by $c^{(a_1a_2)a_3\cdots\,a_n}$ and $c^{[a_1a_2]a_3\cdots\,a_n}$ the symmetrisation and antisymmetrisation, respectively, of the indices (without $\frac1{2}$ factor). We will similarly define
\[
c^{[(a_1a_2)a_3]a_4\cdots\,a_n} = c^{(a_1a_2)a_3a_4\cdots\,a_n} - c^{a_3(a_1a_2)a_4\cdots\,a_n}\,.
\]

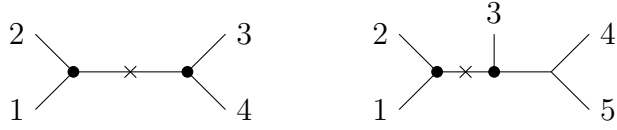
\begin{figure}
    \centering
    \begin{tikzpicture}[baseline=-2.5]
        \draw (-1,-0.5) node[left] {1} -- (-0.5,0);
        \draw (-1,0.5) node[left] {2} -- (-0.5,0);
        \draw (-0.5,0) -- (1,0);
        \draw (1.5,-0.5) node[right] {4} -- (1,0);
        \draw (1.5,0.5) node[right] {3} -- (1,0);
        \filldraw (-0.5,0) circle (2pt);
        \filldraw (1,0) circle (2pt);
        \node[point] at (0.25,0) {};
    \end{tikzpicture} \hspace{2.5em}
    \begin{tikzpicture}[baseline=-2.5]
        \draw (-1,-0.5) node[left] {1} -- (-0.5,0);
        \draw (-1,0.5) node[left] {2} -- (-0.5,0);
        \draw (-0.5,0) -- (1,0);
        \draw (0.25,0.5) node[above] {3} -- (0.25,0);
        \draw (1.5,-0.5) node[right] {5} -- (1,0);
        \draw (1.5,0.5) node[right] {4} -- (1,0);
        \filldraw (-0.5,0) circle (2pt);
        \filldraw (0.25,0) circle (2pt);
        \node[point] at (-0.125,0) {};
    \end{tikzpicture}
    \caption{Diagrammatic representations of eqs.~\eqref{eq:eg4pt} and~\eqref{eq:eg5pt} respectively. The dots represent a squared $X$ and the crosses represent the squared Mandelstam variable. Notice that the one-loop vertex is the full line between the two dots.}
    \label{fig:DiagExample}
\end{figure}

By trial and error, we guessed a set of $m$-point one-loop vertices that lead to the $E$-part of the amplitude (checked numerically up to seven points). The Feynman rules are those given above equation \eqref{eq:4ptCPcheck}, except that the four-point $\hbar^1$-vertex there is substituted by the following $\hbar^1$-vertices of multiplicity four or higher.
\begin{itemize}
\item  $\hbar^1$-vertices $(++\cdots++)$:\footnote{We denote $s_{1\cdots i}=(k_1+k_2+\cdots+k_i)^2$.}
\end{itemize}
\begin{align}
&\sum_{i=2}^{m-2}X(k_1,k_2)^2 \Bigg(\prod_{j=2}^{i-1} \frac{X(k_{1,\cdots,j}\,,k_{j+1})}{s_{1\cdots j}} \Bigg) \frac1{s_{1\cdots i}^2}
\Bigg(\prod_{l=i+1}^{m-2} \frac{X(k_{1,\cdots,l-1}\,,k_{l})}{s_{1\cdots l}} \Bigg) X(k_{m-1},k_m)^2 \nonumber \\
&\quad \cdot 
c^{[[\cdots[(a_1a_2)a_3]\cdots]a_i][a_{i+1}[\cdots[a_{m-2}(a_{m-1}a_m)]\cdots]]}
\,.
\label{eq:SDYM1loopvertices}
\end{align}
It is important to notice that the structure of the colour factors reflects that of the kinematic numerators.
As an illustration, the four-point vertex is simply
\[
\label{eq:eg4pt}
X(k_1,k_2)^2\,\frac1{s_{12}^2}\,X(k_3,k_4)^2\,c^{(a_1a_2)(a_3a_4)}\,,
\]
and a contribution of this vertex at five points takes the form
\begin{align}
&X(k_1,k_2)^2\,\frac1{s_{12}^2}\,X(k_3,k_4+k_5)^2\,c^{(a_1a_2)(a_3b)}\cdot\frac{X(k_4,k_5)}{s_{45}}\,f^{ba_4a_5} 
\nonumber \\
&= X(k_1,k_2)^2\,\frac1{s_{12}^2}\,X(k_3,k_4+k_5)^2\,\frac{X(k_4,k_5)}{s_{45}}\,c^{(a_1a_2)(a_3[a_4a_5])}\,.
\label{eq:eg5pt}
\end{align}
It is helpful to introduce the diagrammatic notation where \eqref{eq:eg4pt} and \eqref{eq:eg5pt} are represented, respectively, by the diagrams in figure~\ref{fig:DiagExample}. By giving the one-loop vertices a representation as a dressed line, we are able to indicate with a dot and a cross the squaring of $X$ and the propagator, respectively.
For instance, the colour-ordered $E$-part of the six-point amplitude is
\begin{align}
      &  {\mathcal A}^{(1)}_{6\;\text{SDYM}} \Big|_{E-\text{part}} =
        \left(\begin{tikzpicture}[baseline=-2.5]
            \draw (-1,-0.5) node[left] {1} -- (-0.5,0);
            \draw (-1,0.5) node[left] {2} -- (-0.5,0);
            \draw (-0.5,0) -- (1,0);
            \draw (0,0.5) node[above] {3} -- (0,0);
            \draw (0.5,0.5) node[above] {4} -- (0.5,0);
            \draw (1.5,-0.5) node[right] {6} -- (1,0);
            \draw (1.5,0.5) node[right] {5} -- (1,0);
            \filldraw (-0.5,0) circle (2pt);
            \filldraw (1,0) circle (2pt);
            \node[point] at (0.75,0) {};
        \end{tikzpicture}\right)
        +
        \frac1{2}\left(\begin{tikzpicture}[baseline=-2.5]
            \draw (-1,-0.5) node[left] {1} -- (-0.5,0);
            \draw (-1,0.5) node[left] {2} -- (-0.5,0);
            \draw (-0.5,0) -- (1,0);
            \draw (0,0.5) node[above] {3} -- (0,0);
            \draw (0.5,0.5) node[above] {4} -- (0.5,0);
            \draw (1.5,-0.5) node[right] {6} -- (1,0);
            \draw (1.5,0.5) node[right] {5} -- (1,0);
            \filldraw (-0.5,0) circle (2pt);
            \filldraw (1,0) circle (2pt);
            \node[point] at (0.25,0) {};
        \end{tikzpicture}\right)\nonumber\\
        &+
        \left(\begin{tikzpicture}[baseline=-2.5]
            \draw (-1,-0.5) node[left] {1} -- (-0.5,0);
            \draw (-1,0.5) node[left] {2} -- (-0.5,0);
            \draw (-0.5,0) -- (1,0);
            \draw (0,0.5) node[above] {3} -- (0,0);
            \draw (0.5,0.5) node[above] {4} -- (0.5,0);
            \draw (1.5,-0.5) node[right] {6} -- (1,0);
            \draw (1.5,0.5) node[right] {5} -- (1,0);
            \filldraw (0,0) circle (2pt);
            \filldraw (1,0) circle (2pt);
            \node[point] at (0.75,0) {};
        \end{tikzpicture}\right)
        +
        \left(\begin{tikzpicture}[baseline=-2.5]
            \draw (-1,-0.5) node[left] {1} -- (-0.5,0);
            \draw (-1,0.5) node[left] {2} -- (-0.5,0);
            \draw (-0.5,0) -- (1,0);
            \draw (0,0.5) node[above] {3} -- (0,0);
            \draw (0.5,0.5) node[above] {4} -- (0.5,0);
            \draw (1.5,-0.5) node[right] {6} -- (1,0);
            \draw (1.5,0.5) node[right] {5} -- (1,0);
            \filldraw (0,0) circle (2pt);
            \filldraw (1,0) circle (2pt);
            \node[point] at (0.25,0) {};
        \end{tikzpicture}\right)
        +
        \frac1{2}\left(\begin{tikzpicture}[baseline=-2.5]
            \draw (-1,-0.5) node[left] {1} -- (-0.5,0);
            \draw (-1,0.5) node[left] {2} -- (-0.5,0);
            \draw (-0.5,0) -- (1,0);
            \draw (0.25,0.4) -- (0.25,0);
            \draw (0.25,0.4) -- node[left] {3} (0,0.8);
            \draw (0.25,0.4) -- node[right] {4} (0.5,0.8);
            \draw (1.5,-0.5) node[right] {6} -- (1,0);
            \draw (1.5,0.5) node[right] {5} -- (1,0);
            \filldraw (-0.5,0) circle (2pt);
            \filldraw (1,0) circle (2pt);
            \node[point] at (0.6,0) {};
        \end{tikzpicture}\right)\nonumber\\
        &+
        \left(\begin{tikzpicture}[baseline=-2.5]
            \draw (-1,-0.5) node[left] {1} -- (-0.5,0);
            \draw (-1,0.5) node[left] {2} -- (-0.5,0);
            \draw (-0.5,0) -- (1,0);
            \draw (0,0.5) node[above] {3} -- (0,0);
            \draw (0.5,0.5) node[above] {4} -- (0.5,0);
            \draw (1.5,-0.5) node[right] {6} -- (1,0);
            \draw (1.5,0.5) node[right] {5} -- (1,0);
            \filldraw (0.5,0) circle (2pt);
            \filldraw (1,0) circle (2pt);
            \node[point] at (0.75,0) {};
        \end{tikzpicture}\right)
        +
         \frac1{2}\left(\begin{tikzpicture}[baseline=-2.5]
            \draw (-1,-0.5) node[left] {1} -- (-0.5,0);
            \draw (-1,0.5) node[left] {2} -- (-0.5,0);
            \draw (-0.5,0) -- (1,0);
            \draw (0,0.5) node[above] {3} -- (0,0);
            \draw (0.5,0.5) node[above] {4} -- (0.5,0);
            \draw (1.5,-0.5) node[right] {6} -- (1,0);
            \draw (1.5,0.5) node[right] {5} -- (1,0);
            \filldraw (0.5,0) circle (2pt);
            \filldraw (0,0) circle (2pt);
            \node[point] at (0.25,0) {};
        \end{tikzpicture}\right)
        +
         \frac1{4}\left(\begin{tikzpicture}[baseline=-2.5]
            \draw (-1,-0.5) node[left] {1} -- (-0.5,0);
            \draw (-1,0.5) node[left] {2} -- (-0.5,0);
            \draw (-0.5,0) -- (1,0);
            \draw (0.25,0.4) -- (0.25,0);
            \draw (0.25,0.4) -- node[left] {3} (0,0.8);
            \draw (0.25,0.4) -- node[right] {4} (0.5,0.8);
            \draw (1.5,-0.5) node[right] {6} -- (1,0);
            \draw (1.5,0.5) node[right] {5} -- (1,0);
            \filldraw (0.25,0) circle (2pt);
            \filldraw (1,0) circle (2pt);
            \node[point] at (0.6,0) {};
        \end{tikzpicture}\right) \nonumber \\
        & + \text{permutations}\{123456\}\,.
        \label{eq:SDYM6pts}
    \end{align}
The numerical coefficients are symmetry factors compensating the overcounting. The first, second and third lines have, respectively, one-loop vertices of multiplicity six, five and four.

Comparing to the action \eqref{eq:qcSDYM'}, the vertices \eqref{eq:SDYM1loopvertices} follow from substituting the last term in the integrand by
\begin{align}
\Bigg( (\Psi \!\Pd\!{}^2\, \Psi )
\, \frac1{\mathbb{1} \cev \square-\cev {\text{ad}}_{\Psi\Pd}}\Bigg)^{\!\! bc}
\Bigg(\, \frac1{\mathbb{1} \vec \square-\vec {\text{ad}}_{\Psi\Pd}} 
\,(\Psi \!\Pd\!{}^2\, \Psi ) \Bigg)^{\!\! cb} 
\label{eq:V1loopE}
\end{align}
times the coupling $a$.
The arrows notation indicates the action of the derivatives. We define\footnote{This takes the form of the (inverse) action of the double covariant derivative $D^\mu D_\mu$ with gauge field \eqref{eq:Au}-\eqref{eq:resAPhi}. 
}
\begin{align}
&\Bigg(\, \frac1{\mathbb{1} \vec \square-\vec {\text{ad}}_{\Psi\Pd}} \; \,(\Psi^{a_1} \!\Pd\!{}^2\, \Psi^{a_2} )\, T^{a_1}T^{a_2} \Bigg)^{\!\! bc} = 
\frac1{\vec \square}\,
f^{b(a_1|e}f^{e|a_2)c}\,(\Psi^{a_1} \!\Pd\!{}^2\, \Psi^{a_2} ) \nonumber \\
&\qquad\qquad\qquad +
(f^{ba_3d}f^{d(a_1|e}f^{e|a_2)c}-f^{b(a_1|e}f^{e|a_2)d}f^{da_3c})\, \frac1{\vec \square}(\Psi^{a_3}  \!\Pd\! \,\frac1{\vec \square}\,(\Psi^{a_1} \!\Pd\!{}^2\, \Psi^{a_2} )) \nonumber \\
&\qquad\qquad\qquad + {\mathcal O}\Big(ffff\,\frac1{ \square}(\Psi P\frac1{ \square}(\Psi P\frac1{ \square}(\Psi P^2\Psi)))\Big) \,,
\end{align}
where we indicated schematically the next term in the geometric series. Focusing on the second line, notice that the colour structure will contribute towards a colour factor $c^{\cdots [a_3(a_1a_2)]}$, while the $ \!\Pd\!$ derivatives will contribute towards a kinematic numerator $\cdots X(k_3,k_1+k_2)X(k_1,k_2)^2$, replicating the form of the vertices \eqref{eq:SDYM1loopvertices}. The squared Mandelstam variable in the denominator of the vertices corresponds to $\square^{-2}$, arising from the overall ${\cev\square}{}^{-1}$ and ${\vec\square}{}^{-1}$ in the left and right factors of \eqref{eq:V1loopE}, respectively.

We did not find it as easy to obtain the vertices leading to the $O$-part of the amplitude. Notice that there is no such vertex at four points due to momentum conservation, while at five points a valid vertex can be obtained by taking the $O$-part of the amplitude off-shell:
\begin{align}
\frac{X(k_1,k_2)}{s_{12}}\frac{X(k_2,k_3)}{s_{23}}\frac{X(k_3,k_4)}{s_{34}}\frac{X(k_4,k_5)}{s_{45}}\frac{X(k_5,k_1)}{s_{51}}\,\varepsilon(1,2,3,4)\; c^{a_1a_2a_3a_4a_5}\,.
\end{align}
One may obtain the six-point vertex in a similar manner, by taking the $O$-part of the amplitude off-shell\footnote{The cyclic product of factors $1/\langle ij\rangle$ is continued off-shell into the cylic product of factors $X(k_i,k_j)/s_{ij}$, times the external state factors that eliminate the $|\eta\rangle$ dependence.} and subtracting the contribution from the five-point vertex dressed with a tree. Higher-point vertices can be obtained similarly. However, we did not find a nice formula for these vertices, so the problem remains of obtaining a decent closed-form expression for the complete SU($N$) quantum-corrected action. We take comfort from the fact that the $E$-part of the SDYM amplitude turns out to be the one that connects to SDG via the double copy.

\subsection{Quantum Feynman rules for SDG from the double copy}

Now we consider the quantum-corrected action for SDG. This follows from the work \cite{Bittleston:2022nfr}, which provided the gravity counterpart to the twistor construction of \cite{Costello:2021bah,Costello:2022wso}. Here, however, we will follow a different route: the double copy. We have reviewed the very simple tree-level story $\text{SDG}\sim(\text{SDYM})^2$ in equation \eqref{eq:3ptvert}, following \cite{Monteiro:2011pc}. This is an off-shell story, which extends to the loop-integrand of the one-loop amplitudes \cite{Boels:2013bi}. However, there is no indication there, or anywhere in the vast literature on one-loop colour-kinematics duality (e.g.~\cite{Bjerrum-Bohr:2013iza,Bern:2013yya,Mafra:2014gja,Geyer:2015bja,Geyer:2015jch,He:2015wgf,He:2016mzd,Bern:2017yxu,He:2017spx,Mafra:2017ioj,Mafra:2018qqe,Edison:2020uzf,Bridges:2021ebs,Porkert:2022efy,Edison:2022jln}), that such a prescription extends in some form to the final expression for the amplitude, i.e.~after loop integration. This is the surprise that we will discuss here, which we may speculate to be part of a bigger story, beyond SDG, even though loop amplitudes are generically not rational.

We recall that we seek the action
\[
\label{eq:qSDG}
S_\text{q.c.SDG}(\phi,\bar\phi) = \int d^4 x \; \Big(\,\bar\phi \big(\square \phi - \{\partial_u \phi,\partial_w \phi\}\big) + V_{\text{1-loop}}[\phi]\,\Big)\,.
\] 
Analogously to the SDYM case, we consider the following Feynman rules.
\begin{itemize}
\item Propagator $(+-)$:\, $\frac{1}{k^2}$\,.
\item $\hbar^0$-vertex $(++-)$:\, $X(k_1,k_2)^2=\langle \eta | k_1k_2 | \eta\rangle^2$\,.
\item External state factors: $\langle \eta i \rangle^{\mp 4}$ for a $\pm$-helicity graviton.
\end{itemize}
To these, we add the vertices arising from $V_{\text{1-loop}}[\phi]$. Our double-copy prescription, based on the tree-level story, is simply to square the numerators of the SDYM vertices \eqref{eq:SDYM1loopvertices}, while leaving the denominators untouched:\footnote{In terms of the amplitudes, the external state factors are also squared from SDYM to SDG, so the full `numerators' of the diagrams are squared.} 
\begin{itemize}
\item  $\hbar^1$-vertices $(++\cdots++)$:
\end{itemize}
\[
\sum_{i=2}^{m-2}X(k_1,k_2)^4 \Bigg(\prod_{j=2}^{i-1} \frac{X(k_{1,\cdots,j}\,,k_{j+1})^2}{s_{1\cdots j}} \Bigg) \frac1{s_{1\cdots i}^2}
\Bigg(\prod_{l=i+1}^{m-2} \frac{X(k_{1,\cdots,l-1}\,,k_{l})^2}{s_{1\cdots l}} \Bigg) X(k_{m-1},k_m)^4 \,.
\]
This leads to expressions that, after including the external state factors, have vanishing weight in the reference spinor $|\eta\rangle$; the complete on-shell cancellation of the amplitudes dependence on $|\eta\rangle$ is not trivial. These vertices can be straightforwardly encoded as
\[
V_{\text{1-loop}}[\phi]=
-4b\Big(\, \frac1{ \square_\phi}\, (\phi \Pd\!{}^4\, \phi )\,\Big)^2
= -4b\Bigg(\, \frac1{ \square -(\phi  \Pd\!{}^2\;\cdot) }\, (\phi \Pd\!{}^4\, \phi )\,\Bigg)^2
\,,
\]
where $\square_\phi$ is the wave operator on the background \eqref{eq:reshphi}, and $b$ is a constant setting the normalisation of the one-loop amplitudes. We have verified that this action reproduces the SDG one-loop amplitudes constructed in \cite{Bern:1998xc}, by making numerical checks up to six points.\footnote{For clarity, our procedure is to take the $E$-part of SDYM, e.g.~\eqref{eq:SDYM6pts} at 6 points, and apply the double-copy prescription for each diagram to obtain the SDG gravity result.} Perhaps surprisingly, the SDG action is simpler than the SDYM one, where only the $E$-part of the vertices arises from \eqref{eq:V1loopE}.

We have connected here the $E$-part of the SDYM amplitude to the SDG amplitude. In fact, a similarity between these objects, when written in a particular way, had already been pointed out in \cite{Bern:1998xc}. We are showing here that this similarity can be written as a double copy. It is not clear to us at present why the $O$-part of the SDYM amplitude, which contains $\varepsilon_{\mu\nu\rho\lambda}$, plays no role in the relation between SDYM and SDG. This may be an important clue to a broader understanding of this type of loop-integrated-level double copy.

In analogy with the SDYM case, we can write the quantum-corrected action in a covariant form as
\[
S_{\text{q.c.SDG}} = S_{\text{SDG}} +  b \int d^4 x \sqrt{|g|} \Bigg(\, \frac1{\square_g}  \Bigg( \frac{\varepsilon^{\rho\lambda\sigma\omega}}{\sqrt{|g|}}\,R^\mu{}_{\nu\rho\lambda} \,R^\nu{}_{\mu\sigma\omega}\Bigg) \Bigg)^2 \,,
\]
where $\square_g$ is the wave operator on the curved background with metric $g_{\mu\nu}$. This form of the action is easily seen to follow from the anomaly-free theory obtained in \cite{Bittleston:2022nfr} via a twistor construction, after integrating out the `axion' $\rho$ and flipping the sign of the resulting interaction term:
\[
S_{\rho-\text{SDG}} = S_{\text{SDG}} + \int  \left( d^4 x \sqrt{|g|}\,\frac1{2}(\square_g\,\rho)^2 + \tilde b\,\rho\, R^\mu{}_\nu\wedge R^\nu{}_\mu \right) \,.
\]
Here, \,$b=\frac1{64}\,\tilde b^2$\, due to the sign flip and the extra $\frac1{2}$ factor discussed at the beginning of section~\ref{sec:explicit}, and note also the two $\frac1{2}$ factors from the definition of the curvature 2-form. Therefore, the result of \cite{Bittleston:2022nfr} already encodes the quantum-corrected action of SDG. We arrived at the same answer in a different way, based on the double copy.


\section{Conclusion}
\label{sec:conclusion}

We have investigated the form of the quantum-corrected actions of SDYM and SDG, focusing on light-cone gauge and building on recent work on the twistor formulation of these theories \cite{Costello:2021bah,Costello:2022wso,Bittleston:2022nfr}. We discussed how the classical integrability currents can be defined in the quantum theory, such that they exhibit the anomaly suggested by W.~Bardeen. Finally, we discovered an unexpected manifestation of the double copy at the level of the one-loop effective vertices of the quantum-corrected actions.

It would be interesting to investigate how the anomaly affects the full (non-self-dual) theories. One interesting question is how the anomaly relates to rational parts of the amplitudes more generally; see e.g.~\cite{Badger:2008cm}. Another question concerns gravity in particular: the two-loop divergence of pure gravity was connected in \cite{Bern:2017puu} to the one-loop same-helicity (e.g.~all-plus) amplitudes. This raises the question of what happens if we subtract the anomaly explicitly, by considering an action
\[
S=S_{\text{Einstein-Hilbert}} -  b \int d^4 x \sqrt{|g|} \Bigg(\, \frac1{\square_g}  \Bigg( \frac{\varepsilon^{\rho\lambda\sigma\omega}}{\sqrt{|g|}}\,R^\mu{}_{\nu\rho\lambda} \,R^\nu{}_{\mu\sigma\omega}\Bigg) \Bigg)^2 \,,
\]
where $b\sim\hbar$ is a constant fixed by the normalisation of the one-loop amplitudes. In our previous discussion, (minus) the added term is identified as the quantum correction to the SDG action. Here, instead, it is added to the ordinary Einstein-Hilbert action in order to eliminate same-helicity one-loop amplitudes. One may speculate that, if the findings of \cite{Bern:2017puu} happen to generalise so that all the ultraviolet divergences of pure gravity are related to the anomaly we discussed, then eliminating this anomaly could eliminate the divergences. In fact (as often emphasised by Z.~Bern), the only two explicitly known ultraviolet divergences in four-dimensional pure (super)gravity are associated to anomalies. These are the two-loop divergence in pure gravity we mentioned, and the four-loop divergence in ${\mathcal N}=4$ supergravity studied in \cite{Carrasco:2013ypa,Bern:2013uka,Bern:2017rjw,Bern:2019isl,Carrasco:2022lbm}.

Setting a lower bar of ambition, it would be nice to connect the anomaly explicitly to known properties of the amplitudes in the self-dual theories. One example is the one-loop `dimension-shifting' formula \cite{Bern:1996ja,Britto:2020crg} relating SDYM to the MHV sector in maximally supersymmetric Yang-Mills theory, a theory that is thought to be quantum-integrable (in the planar limit, which at one loop determines also the non-planar contribution). In \cite{He:2015wgf}, similar structures based on the self-dual kinematic algebra were seen to arise in the two theories. Another example is the conformal invariance of the SDYM amplitudes demonstrated in \cite{Henn:2019mvc}, which should find a natural explanation in the twistor approach. Another direction is to consider amplitudes in related theories, e.g.~self-dual Einstein-Yang-Mills amplitudes \cite{Nandan:2018ody,Faller:2018vdz,Porkert:2022efy} or chiral higher-spin theories \cite{Ponomarev:2016lrm,Ponomarev:2017nrr,Krasnov:2021nsq,Tran:2021ukl,Skvortsov:2022syz,Sharapov:2022faa,Tran:2022tft,Monteiro:2022lwm,Herfray:2022prf,Adamo:2022lah}.

Our most surprising finding is the appearance of an explicit double copy after loop integration. One question is whether there is a gauge-invariant version of this double copy. There exist studies of related structures, e.g.~\cite{Bjerrum-Bohr:2011jrh}, which may benefit from our results regarding the $E$/$O$ splitting of the SDYM amplitudes. Naturally, the loop amplitudes we considered here are rational functions, so they are very special. The notion of double copy we employed is similar to the tree-level story, and does not straightforwardly extend to functions with branch cuts, which are the generic case. It is difficult to say at this stage whether our finding is the first evidence of a more general story.

\section*{Acknowledgements}

We thank Zvi Bern, Andi Brandhuber, Graham Brown, Josh Gowdy, Lionel Mason, Natalie Paquette, Atul Sharma, Gabriele Travaglini and Chris White for discussions and comments. RM is supported by the Royal Society via a University Research Fellowship. RSM was supported by the Royal Society via a studentship grant. SW's studentship is supported by STFC.

\appendix
\section{Brute-force one-loop vertices}
\label{sec:bruteforce}

In this appendix we briefly illustrate the brute-force approach to computing the one-loop effective $m$-point vertices in SDYM and SDG. For each $m$, this involves computing the one-loop diagram with $m$ off-shell external legs directly attached to the loop, using the light-cone gauge Feynman rules in eq.~\eqref{eq:3ptvert}, by explicitly performing the loop integration in a given regularisation scheme. Here, we demonstrate this procedure for the $m = 2,3,4$ colour-ordered vertices in dimensional regularisation, using the ``X'' Mathematica package to perform the loop integrals.

\subsection{$m=2$}

At two points, the relevant diagram is the bubble, given by
    \begin{equation}\label{eq:I2main}
        V^{(2)}_{\text{SDYM}} = \mu^{2\epsilon}\int \frac{d^D l}{(2\pi)^D}\frac{X(l,k)X(l+k,-k)}{l^2(k+l^2)}
        = -\mu^{2\epsilon}\int \frac{d^D l}{(2\pi)^D}\frac{X(l,k)^2}{l^2(k+l^2)}.
    \end{equation}
Here $k$ and $l$ are the external and loop momenta respectively, and $\mu$ is the regularisation scale. The loop momenta is $(4-2\epsilon)$-dimensional. We now employ some standard techniques. Firstly, we introduce the Feynman parametrisation,
    \begin{equation}
        \frac{1}{l^2(k+l^2)} = \int_0^1 dx \frac{1}{\left[l^2+x((k+1)-l^2)\right]^2}
        = \int_0^1 dx \frac{1}{\left[(l+xk)^2 - \Delta\right]^2},
    \end{equation}
where $\Delta = -xk^2(1-x)$. Then, we perform the shift $l \to \tilde{l} = l + xk$, such that eq.~\eqref{eq:I2main} becomes
    \begin{align}
        V^{(2)}_{\text{SDYM}} &= -\mu^{2\epsilon}\int_0^1 dx \int \frac{d^D \tilde{l}}{(2\pi)^D}\frac{X(\tilde{l},k)^2}{(\tilde{l}^2-\Delta)^2} \\
        &= -\langle\eta|\sigma^{\mu}k|\eta\rangle\langle\eta|\sigma^{\nu}k|\eta\rangle \mu^{2\epsilon}\int_0^1 dx \int \frac{d^D \tilde{l}}{(2\pi)^D}\frac{\tilde{l}_{\mu}\tilde{l}_{\nu}}{(\tilde{l}^2-\Delta)^2}.
    \end{align}
The denominator is symmetric in $\tilde{l}$, and thus we can make the replacement
    \begin{equation}
        \tilde{l}_{\mu}\tilde{l}_{\nu} \to \frac{1}{D}\tilde{l}^2\eta_{\mu\nu}.
    \end{equation}
The metric contracts with the indices in the prefactor, and  by applying the identity $\sigma^{\mu}_{\alpha\dot{\alpha}}\sigma_{\mu\beta\dot{\beta}} = 2\epsilon_{\alpha\beta}\epsilon_{\dot{\alpha}\dot{\beta}}$, we find that
    \begin{equation}
        \langle\eta|\sigma^{\mu}k|\eta\rangle\langle\eta|\sigma_{\mu}k|\eta\rangle \propto X(k,k) = 0.
    \end{equation}
We see, therefore, that $V^{(2)}_{\text{SDYM}} = 0$ even off-shell. There is a subtle issue of regularisation (see e.g.~\cite{Chattopadhyay:2020oxe,Chattopadhyay:2021udc}), but our choice here is consistent with the following higher-point computations. 

A similar calculation can be carried out in SDG, where the integrand is constructed following the double copy prescription discussed in~\cite{Boels:2013bi}. We find, similarly, that $V^{(2)}_{\text{SDG}} = 0$ off-shell.

\subsection{$m=3$}

At three points, the relevant diagram is the triangle, given by
    \begin{equation}\label{eq:I3main}
        V^{(3)}_{\text{SDYM}} = \mu^{2\epsilon}\int \frac{d^D l}{(2\pi)^D} \frac{X(l,1)X(l,2)X(l+2,3)}{l^2(l-k_1)^2(l+k_2)^2},
    \end{equation}
    where $X(l,1)=X(l,k_1)$, etc.
Due to the linearity of $X(k,k')$ in its arguments this reduces to a sum of a rank-3 and a rank-2 tensor integrals. These integrals can be evaluated with Passarino-Veltman reductions using the ``X'' Mathematica package, with the result
    \begin{equation}\label{eq:I3integrated}
        V^{(3)}_{\text{SDYM}} = \frac{i}{16 \pi^2} X(2,3)^3\left[
            \sum_{i=1}^3 a_i \ln\left(-\frac{4\pi\mu^2}{k_i^2}\right)
            + b S_{c_0}(k_1^2,k_2^2,k_3^2)
            + \mathcal{R}
        \right],
    \end{equation}
where the coefficients are
    \begin{align}
        a_1 &= -\frac{k_1^2}{3\lambda}\left[
            (k_2^2+k_3^2)(k_2^2+k_3^2-k_1^2)^3 + 4k_2^2k_3^2(k_2^4+k_3^4-k_1^4)
        \right.\nonumber \\ &\left.\hspace{4em}
            + 18k_1^2k_2^2k_3^2(k_2^2+k_3^2-k_1^2) - 24k_2^3k_3^4
        \right], \\
        a_2 &= a_1|_{k_1 \leftrightarrow k_2}, \\
        a_3 &= a_1|_{k_1 \leftrightarrow k_3}, \\
        b &= \frac{2k_1^2k_2^2k_3^2}{\lambda^3}\left[
            k_1^4(k_2^2+k_3^2-k_1^2) + k_2^4(k_3^2+k_1^2-k_2^2) + k_3^4(k_1^2+k_2^2-k_3^2) + 4k_1^2k_2^2k_3^2
        \right].
    \end{align}
Here, $\lambda = \lambda(k_1^2,k_2^2,k_3^2)$ is the K\"{a}ll\'{e}n function
    \begin{equation}
        \lambda = 2(k_1^4+k_2^4+k_3^4)-(k_1^2+k_2^2+k_3^2)^2.
    \end{equation}
Finally, we have the rational part
    \begin{equation}
        \mathcal{R} = \frac{1}{6\lambda^2}\left[
            k_1^4(k_2^2+k_3^2-k_1^2) + k_2^4(k_3^2+k_1^2-k_2^2) + k_3^4(k_1^2+k_2^2-k_3^2) + 14k_1^2k_2^2k_3^2
        \right],
    \end{equation}
and the scalar function
    \begin{align}
        S_{c_0}(k_1^2,k_2^2,k_3^2) &= \frac{1}{\sqrt{\lambda}}\sum_{\text{cyc}(1,2,3)}\left[
            \text{Li}_2\left(\frac{k_1^2+k_2^2-k_3^2+\sqrt{\lambda}}{k_1^2+k_2^2-k_3^2-\sqrt{\lambda}}\right)
            \right. \nonumber\\&\left.\hspace{7em}
            - \text{Li}_2\left(\frac{k_1^2+k_2^2-k_3^2-\sqrt{\lambda}}{k_1^2+k_2^2-k_3^2+\sqrt{\lambda}}\right)
        \right],
    \end{align}
where $\text{Li}_2$ is a dilogarithm and the sum is over cyclic permutations.

In the limit where two legs are taken on-shell, only the rational part survives and we are left with
    \begin{equation}
        \label{ap:V3}
        V^{(3)}_{\text{SDYM}}\big|_{k_2^2,k_3^2 \to 0} = \frac{i}{96\pi^2}\frac{X(2,3)^3}{k_1^2}.
    \end{equation}
When the final leg is taken on-shell this vanishes due to 3-point massless kinematics. This is as expected, since the 3-point all-plus amplitude vanishes. 

The expression \eqref{eq:I3main} for the 3-point one-loop vertex is not exactly appealing.  It turns out that keeping only the rational part of the vertex, proportional to $\mathcal{R}$, is sufficient, as we checked explicitly by computing the 3-point vertex contribution to the 4-point amplitude. While this is an improvement, we should note that the 3-point amplitude vanishes, so it is unhelpful to use a formalism where a vertex exists at this order. This is the reason why in the main body of the paper we used the vertices associated to the twistor constructions, which appear firstly at 4 points.

The same computation can be done in SDG, where we square the numerator in eq.~\eqref{eq:I3main}. This yields rank-four, -five, and -six tensor integrals, which can be evaluated in Mathematica. The result takes a similar form to the SDYM case in eq.~\eqref{eq:I3integrated}, but with more complicated coefficients and an overall factor of $X(2,3)^6$. Taking two of the legs on-shell, only the rational part survives, and we obtain
    \begin{equation}
        V^{(3)}_{\text{SDG}}\big|_{k_2^2,k_3^2 \to 0} \propto \frac{X(2,3)^6}{k_1^2},
    \end{equation}
    up to a numerical factor.
This vanishes when the final leg goes on-shell, as expected.

\subsection{$m=4$}

At four points, the computation is significantly more involved. The relevant diagram is the off-shell box.  Once the loop integral is evaluated, the result has a similar structure as at three points, containing a rational part, logarithms, and dilogarithms, albeit with a far greater number of these terms.

We will not write here the resulting vertex, but to illustrate a point we consider instead the colour-ordered `off-shell amplitude' (without external state factors). This is given by the sum of the 4-point one-loop vertex and the four diagrams with the one-loop 3-point vertex attached to the tree-level 3-point vertex. Taking all but one legs on-shell, only the rational parts of the loop vertices contribute, and after some massaging we obtain the following expression
    \begin{align}
        \label{ap:I4}
        I^{(4)}_{\text{SDYM}}\big|_{k_2^2,k_3^2,k_4^2 \to 0}
        &\propto \frac{X(1,2)X(3,4)^3}{s_{34}^2} + \frac{X(2,3)^3X(4,1)}{s_{23}^2} \nonumber\\
        &+ \frac{X(2,3)X(3,4)}{s_{23}s_{34}} \left[X(1,2)X(2,3)+X(3,4)X(4,1)+X(1,2)X(4,1)\right].
    \end{align}
    This collapses in the fully on-shell case to
    \[
    I^{(4)}_{\text{SDYM}}\big|_{k_1^2,k_2^2,k_3^2,k_4^2 \to 0} \propto \frac{X(1,2)X(2,3)X(3,4)X(4,1)}{s_{12}s_{23}}\,,
    \]
    which leads directly to the correct 4-point amplitude.
    The fact that only the rational parts contribute in these limits means that the rational parts can be taken as the vertices at this multiplicity. It is possible that this extends to any multiplicity and that, therefore, this procedure can lead to rational vertices to all orders. However, we have clearly highlighted the disadvantages in this appendix. Such a brute-force approach is even less feasible for SDG.
    
The reason why we included expressions \eqref{ap:V3} and  \eqref{ap:I4} is that these objects with one leg off-shell (and after dividing by $1/k_1^2$) are the one-loop Berends-Giele currents in this approach. These currents are associated to perturbative solutions to the equations of motion. In particular, the one-loop Berends-Giele currents above are one-loop contributions to the perturbative solution for $\bar\Psi$ in the quantum-corrected theory. The quantum part of the solution for $\bar\Psi$ is generated only by $\Psi$ sources, i.e.~positive-helicity sources. (While the field $\bar\Psi$ has negative helicity as a solution, when we think about a diagram the `measured' field is outgoing, while the sources are incoming, so this does correspond to an all-plus diagram when we take all legs to be incoming.) We tried without success to find a closed-form solution to the quantum-corrected equation of motion \eqref{eq:qeomPsib}, which we hoped could have similar features to the known (see \cite{Bardeen:1995gk,Cangemi:1996rx}) perturbative solution to the equation of motion \eqref{eq:qeomPsi}; the latter equation is unchanged from classical to quantum level. Nevertheless, we note that even in this brute-force approach, the one-loop currents are rational functions up to 4 points, a property which we presume extends to higher points (and is likely related to the results of \cite{Blanco:2020akb}).

\bibliography{refs}
\bibliographystyle{JHEP}

\end{document}